\documentclass[twocolumn,showpacs,preprintnumbers,amsmath,amssymb]{revtex4-1}

\usepackage{graphicx}
\usepackage{dcolumn}
\usepackage{bm}
\usepackage[titletoc]{appendix}  
\usepackage{color}
\usepackage{hyperref}

\begin{document}


\title{Fundamental properties of cooperative contagion processes}


\author{Li Chen}
\email{chenl@snnu.edu.cn}
\affiliation{Robert Koch-Institute, Nordufer 20, 13353 Berlin, Germany}
\affiliation{Max Planck Institute for the Physics of Complex Systems, 01187 Dresden, Germay}
\affiliation{School of Physics and Information Technology, Shaanxi Normal University, Xi'an 710062, China}
\author{Fakhteh Ghanbarnejad}
\email{fakhteh@pks.mpg.de}
\affiliation{Robert Koch-Institute, Nordufer 20, 13353 Berlin, Germany}
\affiliation{Max Planck Institute for the Physics of Complex Systems, 01187 Dresden, Germay}
\author{Dirk Brockmann}
\email{dirk.brockmann@hu-berlin.de}
\affiliation{Robert Koch-Institute, Nordufer 20, 13353 Berlin, Germany}
\affiliation{Institute for Theoretical Biology \& Integrative Research Institute for the Life Sciences, Humboldt Universit\"at zu Berlin, Philippstr. 13, Haus 4, 10115 Berlin, Germany}


\begin{abstract}
We investigate the effects of cooperativity between contagion processes that spread and persist in a host population. We propose and analyze a dynamical model in which individuals that are affected by one transmissible agent $A$ exhibit a higher than baseline propensity of being affected by a second agent $B$ and vice versa. The model is a natural extension of the traditional SIS (Susceptible-Infected-Susceptible) model used for modeling single contagion processes. We show that cooperativity changes the dynamics of the system considerably when cooperativity is strong. The system exhibits discontinuous phase transitions not observed in single agent contagion, multi-stability, a separation of the traditional epidemic threshold into different thresholds for inception and extinction as well as hysteresis. These properties are robust and are corroborated by stochastic simulations on lattices and generic network topologies. Finally, we investigate wave propagation and transients in a spatially extended version of the model and show that especially for intermediate values of baseline reproduction ratios the system is characterized by various types of wave-front speeds. The system can exhibit spatially heterogeneous stationary states for some parameters and negative front speeds (receding wave fronts). The two agent model can be employed as a starting point for more complex contagion processes, involving several interacting agents, a model framework particularly suitable for modeling the spread and dynamics of microbiological ecosystems in host populations. 
\end{abstract}

\pacs{05.90.+m, 89.75.Hc, 87.23.Cc}

\maketitle

\section{Introduction}
\begin{figure*}[t!]
\includegraphics[scale=1]{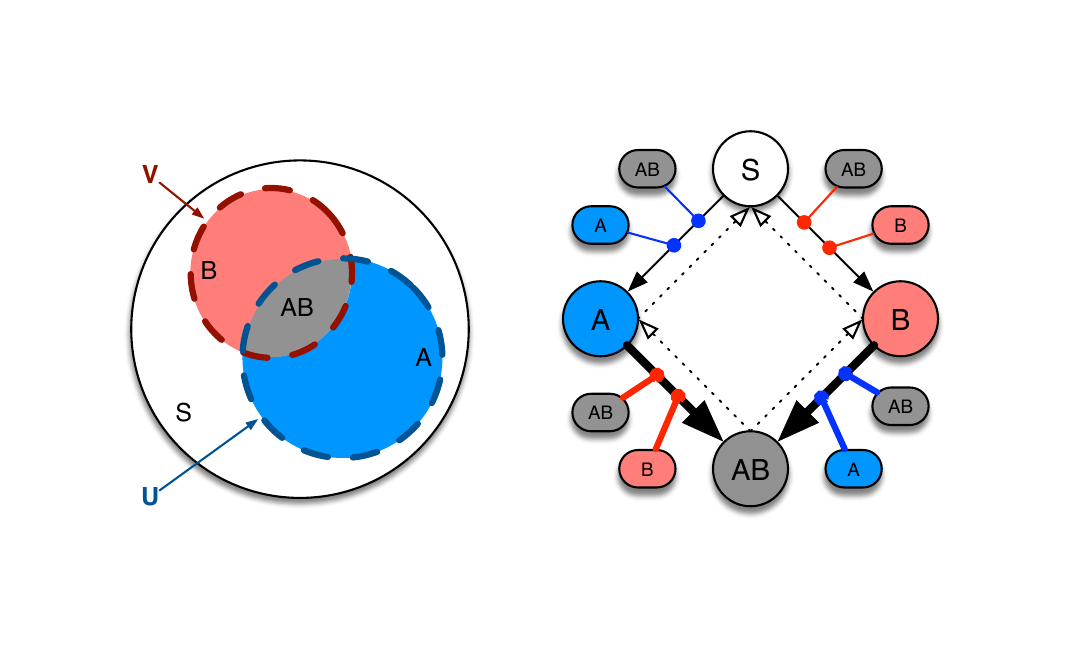}
\caption{\textbf{Cooperativity of two contagion processes}. Two agents, $A$ and $B$, spread in a host population. We classify the state of a host individual by letters $S$ (white), $A$ (blue), $B$ (red), and $AB$ (grey) corresponding to being susceptible, infected only by $A$, infected only by $B$, and infected by both $A$ and $B$, respectively. The state of the population can be defined by the subsets $\mathcal{S}$, $\mathcal{A}^{+}$, $\mathcal{B}^{+}$, $\mathcal{A}^{+}\cap\mathcal{B}^{+}$ corresponding to the sets of susceptibles, infected by $A$ (interior of blue dashed circle), infected by $B$ (interior of red dashed circle), and infected by both (grey area), respectively. Note that the sets $\mathcal{A}^{+}$ and $\mathcal{B}^{+}$ include individuals in state $AB$. The relative size (fraction of individuals) of $\mathcal{A}^{+}$, $\mathcal{B}^{+}$, $\mathcal{A}^{+}\cap\mathcal{B}^{+}$ and $\mathcal{S}$ is denoted by $u$, $v$, $w$, and $1-u-v+w$, respectively. Contagion dynamics is determined by 12 reactions depicted on the right. Susceptibles $S$ acquire $A$ by interacting with individuals from set $\mathcal{A}^{+}$ ($A$ or $AB$ individuals) at rate $\alpha_{\text{A}}$. Likewise, susceptibles $S$ acquire $B$ by interacting with individuals in set $\mathcal{B}^{+}$ ($B$ or $AB$ individuals) at rate $\alpha_{\text{B}}$. Cooperativity means that individuals in state $A$ ($B$) acquire agent $B$ ($A$) at higher rates interacting with individuals in set $\mathcal{B}^{+}$($\mathcal{A}^{+}$) symbolized by the thicker interaction lines in the reaction scheme. Dashed lines symbolize recovery events.
}
\label{Fig:fig1}
\end{figure*}

Contagion processes abound in nature, ranging from the spread of infectious diseases in host populations \cite{anderson1992}, the spread of information in social networks \cite{gruhl2004}, the adaptation of technology and norms \cite{bicchieri2005, rogers2010}, to activation patterns in neural tissue \cite{buzsaki2004, buzsaki2006}, and escape mechanisms from predators in schooling fish \cite{cooper2015}. Dynamical computational models are an essential tool for understanding phenomena in all of these contexts. Their application to the spread of infectious diseases has flourished in recent years \cite{hufnagel2004, colizza2007, fraser2009, brockmann2013}, primarily because of the relevance to human health and the spread of human infectious diseases. Dynamical models cover a broad scope in terms of complexity, ranging from qualitative models that focus on universal features of the observed phenomenon \cite{Kermack1927, Hethcote2000}, network models that account for population structure or host mobility patterns \cite{barrat2008, pastor2001, colizza2006, brockmann2006, gonzalez2008, balcan2009, belik2011}, to sophisticated, large-scale agent-based models \cite{eubank2004, van2011} that incorporate high resolution data on multi-scale transportation, demographics, epidemiological factors, and behavioral response rules. State-of-the-art computational models have become remarkably successful in reproducing observed patterns and predicting the trend of ongoing epidemics. 

Most epidemic models focus on the transmission dynamics of single, symptomatic pathogenic bacteria or viruses because in most applications it can reasonably be assumed that the phenomena are dominated by host pathogen interactions. A variety of infectious diseases exist, however, that interact either directly or indirectly e.g. by altering the susceptibility of the host with respect to infection with another pathogen. Furthermore, transmissions of bacterial microorganism between host individuals is not restricted to species that cause disease. The transmission and spread of commensal or mutualistic bacteria as part of the host's microbiome is generic, in fact also often required to sustain a healthy, host specific microbiome. Especially the transmission of bacterial species of the human microbiome has attracted much attention in very recent studies \cite{sommer2013, coyte2015}. Microbiotic species are part of a complex microbiological ecosystem of a host, with a densely connected set of metabolic connections \cite{patil2015}. It is reasonable to assume that these interactions, and the presence of particular species in a host's microbiome impacts the propensity of colonization with another. In social science, the adoption of a certain behavioral patterns may impact the propensity to adopt another pattern if exposed to it. Therefore, it is important to understand the basic mechanisms and effects that are generated by interactions of contagion processes in general. 

Early network theoretic work focused on competitive coinfection \cite{funk2010, marceau2011, sahneh2014, newman2005, ahn2006, karrer2011, poletto2013, poletto2015} with important applications to infection dynamics of virus strains that induce cross immunity. Multiplex network approaches have also been applied in this context \cite{buldyrev2010, de2013}, in which each contagion process evolves along a different set of links in the same population \cite{funk2010, marceau2011, sahneh2014, sanz2014}. 
\begin{figure*}[t!]
\includegraphics[scale=0.45]{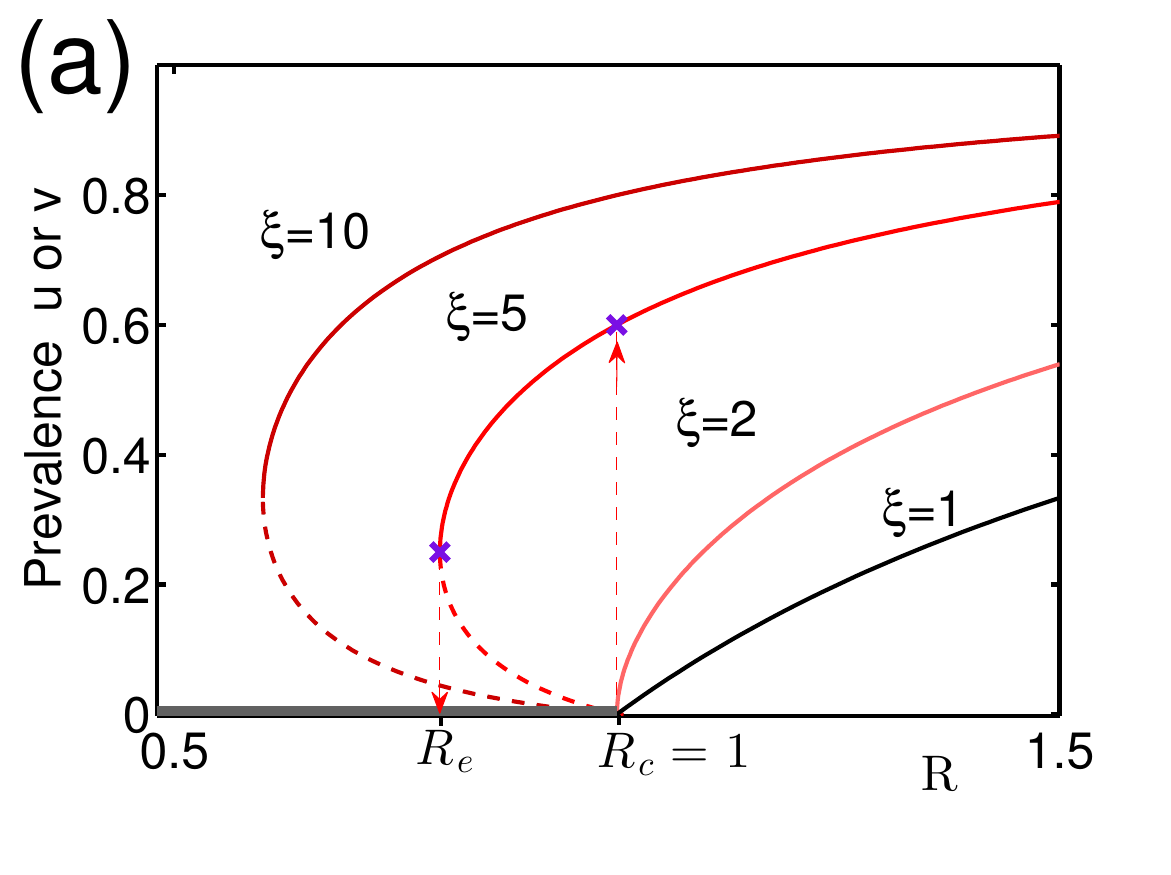}
\includegraphics[scale=0.45]{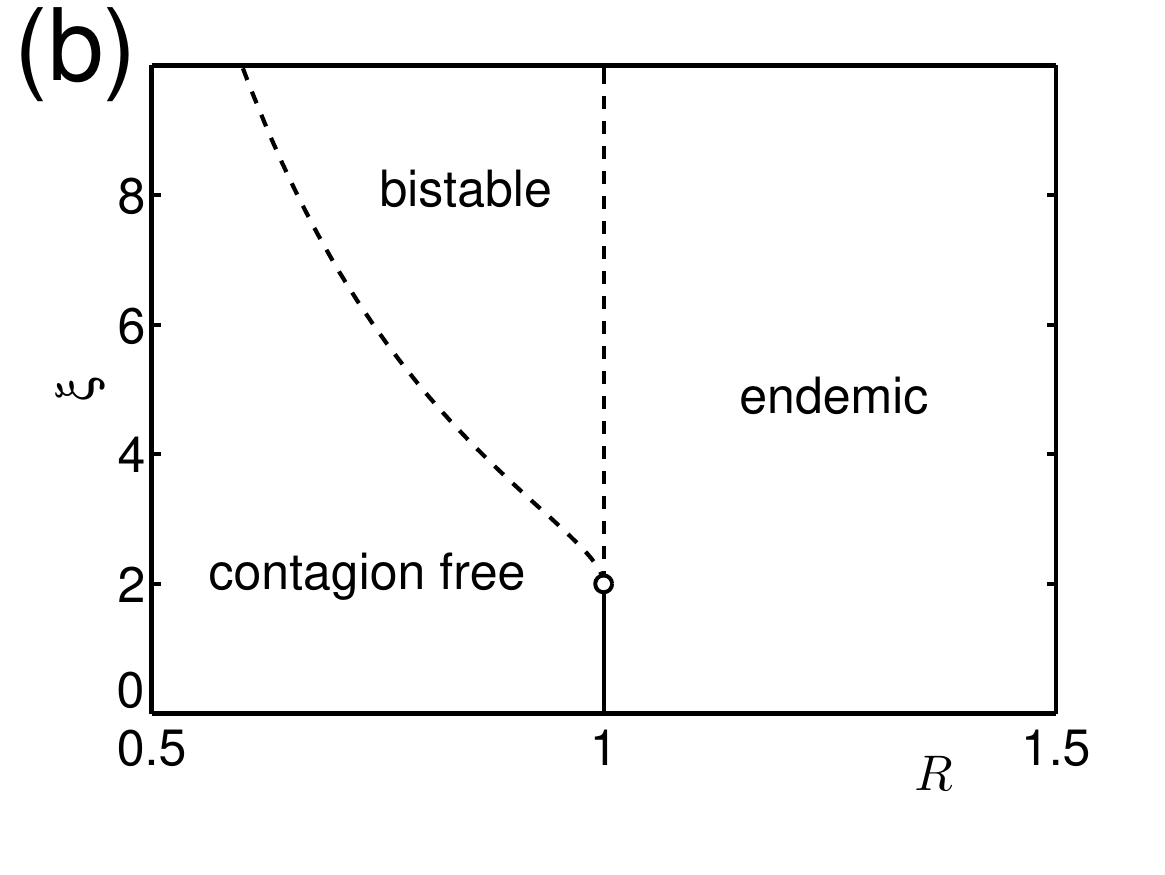}
\caption{\textbf{Bifurcation analysis of cooperative contagion processes}. (a) For various values of the cooperativity coefficient $\xi$ the stationary states of the symmetric system (Eqs. (\ref{eq:symmetricsystem}) are depicted. When $\xi$ is greater than the critical cooperativity $\xi_{c}=2$ a regime $R_{e}<R<R_{c}$ exists in which the system exhibits three stationary states, the stable trivial state $u=v=0$ (grey line), another stable endemic state (upper branch, solid red lines) and an unstable intermediate state (dashed red line), see Eqs. (\ref{eq:knut}). In this regime small perturbation to the $u,v=0$ state will not cause a transition to the endemic branch. Only if perturbations are sufficiently large (crossing the unstable fixed point branch) the system will approach the endemic state. This behavior implies that when subjected to sporadic small perturbations while increasing $R$, the system will remain near the stable contagion free state until the upper critical point $R_{c}=1$ is crossed at which point the system will generate a discontinuous jump, similar to a first order phase transition. The vertical dashed lines illustrate the hysteresis loop. (b) Phase diagram of the system in parameter space, separating three asymptotic states: contagion free, endemic, and bistable, with discontinuous (dashed line) and continuous (solid line) transitions at the interfaces. The circle denotes the tri-critical point at ($R_{c}=1$, $\xi_{c}=2$), which separates the continuous and discontinuous outbreak transitions.
}
\label{Fig:fig2}
\end{figure*}

Only recently, cooperative contagion in which infection with one transmissible agent facilitates infection with another was investigated \cite{chen2013, cai2015, grassberger2016, hebert2015, janssen2016}. These studies mainly focused on transient dynamics of the generic SIR (Susceptible-Infected-Recovery) model in which individuals acquire immunity after infection. In Ref. \cite{cai2015}, a simple SIR coinfection model was investigated within the framework of cooperative bond percolation. This model exhibits avalanche-like outbreak scenarios, depending on the level of cooperation and the structure of the underlying transmission network. Analytical insights were obtained for cooperative bond percolation in multiplex systems \cite{allard2015, azimi2016}. Furthermore, it has been found that highly clustered structures in population aid the proliferation of coinfections, contrary to the effect observed in single disease dynamics \cite{hebert2015}. Because most of these models focus on transient SIR dynamics they can't capture situations in which a steady supply of susceptibles permits the existence of a stable endemic state, such as the SIS or SIRS or SIR model with vital processes. Particularly in these systems, some fundamental issues remain elusive: What basic dynamical features can we expect in cooperative contagion processes? To what extent does cooperativity change the classic outbreak scenario, what is the nature of transitions to endemic states? Can we expect multi-stability and multiple thresholds? How does cooperativity impact spatial propagation? 

Here we introduce and investigate a model for the dynamics of two transmissible, interacting agents (labeled $A$ and $B$). The model is based on the well-known SIS model in which host individuals are either susceptible (S) or infected (I). Susceptibles can be infected with either agent. When infected with say $A$ they can transmit $A$ to other susceptibles. Infecteds remain in the infectious state for a typical period after which they recover and susceptible again. The transmission dynamics of agents $A$ and $B$ are governed by agent specific baseline reproduction numbers $R_{\text{A}}$ and $R_{\text{B}}$, respectively that describe the dynamics of an agent in the absence of the other. We incorporate cooperativity by two additional parameters, the cooperativity coefficients $\xi_{\text{A}}$ and $\xi_{\text{B}}$ that capture influence of an infection with $A$ on the subsequent infection with $B$ and vice versa.

Based on this model, we show that cooperativity between contagion processes generates a variety of interesting properties that are absent in single agent dynamics. For sufficiently strong cooperativity, increasing the baseline reproduction number of one or both agents yields abrupt, discontinuous outbreak transitions and multi-stability (i.e. the coexistence of different stable asymptotic states). Furthermore, cooperativity exhibits dynamical hysteresis, a consequence of the split of the ordinary epidemic threshold into two separate thresholds (an inception and extinction threshold). We derive these features analytically in a deterministic well-mixed model. Their robustness is corroborated by numerical simulations of analogous stochastic dynamical processes on both lattices and generic network systems. Finally we investigate cooperative contagion in spatially extended systems. We show that the interplay of different thresholds and hysteresis yields a rich set of wavefront dynamics and invasion dynamics, e.g. accelerated propagation in certain parameter regimes, stable heterogeneous patterns as well as negative wavefront speeds (receding wavefronts).

\begin{figure*}[t!]
\includegraphics[scale=0.6]{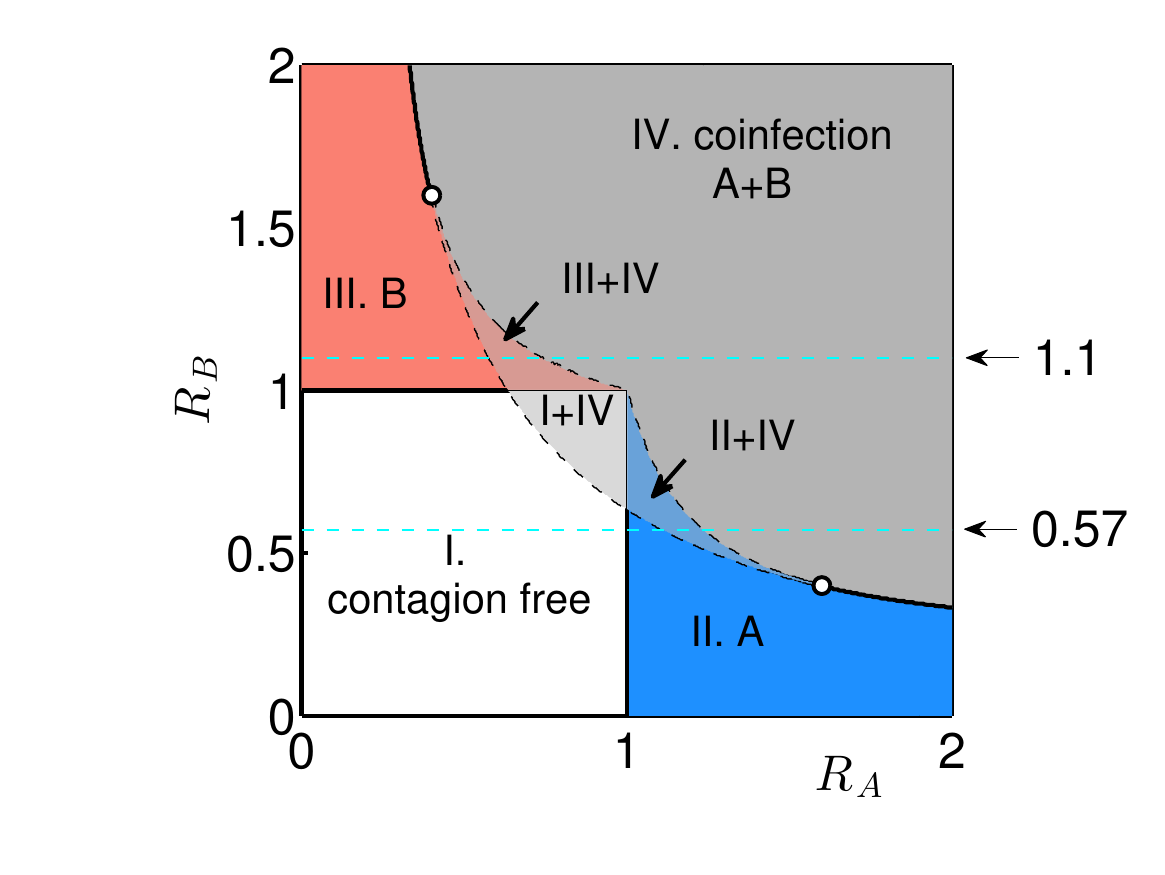}
\caption{\textbf{Phase diagram of generic cooperative contagion for fixed cooperativity coefficient $\xi=5$}. In the parameter space spanned by the baseline reproduction ratio $R_{A}$ and $R_{B}$ we observe four phases in which only a single stable state exists: $S$ (region I), $A$ (region II), $B$ (region III), and $AB$ (region IV). When baseline reproduction is near unity the system exhibits additional regimes characterized by coexisting stable states (bounded by the black dashed lines): coexistence of $S$ and $AB$ (region I+IV), of $A$ and $AB$ (region II+IV), $B$ and $AB$ (region III+IV). Regimes are separated by different types of bifurcations. Solid lines represents the ordinary transcritical bifurcation, dashed lines represent discontinuous transitions. The black circles denote tri-critical points where bifurcation types merge. The two horizontal dashed lines correspond to the panels depicted in Fig. \ref{Fig:cuts}.
}
\label{Fig:phasediagram}
\end{figure*}
\section{Cooperative Contagion }

Our model is an extension of the generic SIS compartmental model:
host individuals are either susceptible (S) or infected (I) and change
state by two reactions, the transmission of the infection $S+I\rightarrow2I$
and recovery $I\rightarrow S$ at rates $\alpha$ and $\beta$, respectively.
In a well mixed, large, and conserved population the fraction of infected
individuals $u(t)$ can be described by $\dot{u}=\alpha u(1-u)-\beta u$.
The basic reproduction ratio is defined by $R=\alpha/\beta$. For
$R<1$ the trivial state $u=0$ is globally stable. If $R$ is increased
beyond the critical threshold $R_{c}=1$ the system exhibits a transcritical
bifurcation, $u=0$ becomes unstable and $u=1-R^{-1}$ is the stable
endemic state in which transmission and recovery events balance. The
SIS system thus exhibits a continuous transition as $R$ crosses the
critical threshold $R_{c}=1$. Analogous stochastic lattice models
in which lattice sites can transmit to neighboring sites and recover
exhibit the same type of threshold behavior and a continuous phase
transition. Here, we consider a generalization of the SIS model that
captures the dynamics of two interacting transmissible agents: $A$
and $B$. A host can be in one of four states $S$, $A$, $B$, and
$AB$, corresponding to susceptible, infected with $A$ but not $B$,
infected with $B$ but not $A$, and infected with both $A$ and $B$,
respectively, see Fig.~\ref{Fig:fig1}. Transmissions in this system
occur by interactions of host individuals in these four different
states and can be summarized as follows 
\begin{eqnarray}
A\vee AB+S\xrightarrow{\alpha_{\text{A}}}A\vee AB+A, \nonumber \\ 
B\vee AB+S\xrightarrow{\alpha_{\text{B}}}B\vee AB+B,\nonumber \\
A\vee AB+B\xrightarrow{\alpha_{\text{BA}}}A\vee AB+AB, \nonumber \\ 
B\vee AB+A\xrightarrow{\alpha_{\text{AB}}}B\vee AB+AB,\label{eq:reactions}
\end{eqnarray}
where e.g. $A\vee AB$ represents an individual that is either in
state $A$ or in state $AB$ such that e.g. the first reaction represents
the transmission of agent $A$ to a susceptible individual. The system
is defined by four different transmission rates $\alpha_{\text{A}}$,
$\alpha_{\text{B}}$, $\alpha_{\text{BA}}$, and $\alpha_{\text{AB}}$
that correspond to transmission of $A$ to a susceptible, of $B$
to a susceptible, of $A$ to an individual carrying $B$, of $B$
to an individual carrying $A$, respectively. For simplicity we assume
uniform recovery rates:
\begin{eqnarray}
AB\xrightarrow{\beta}A\vee B, & \qquad & A\vee B\xrightarrow{\beta}S.\label{eq:recovery}
\end{eqnarray}
\begin{figure*}[t!]
\includegraphics[scale=0.45]{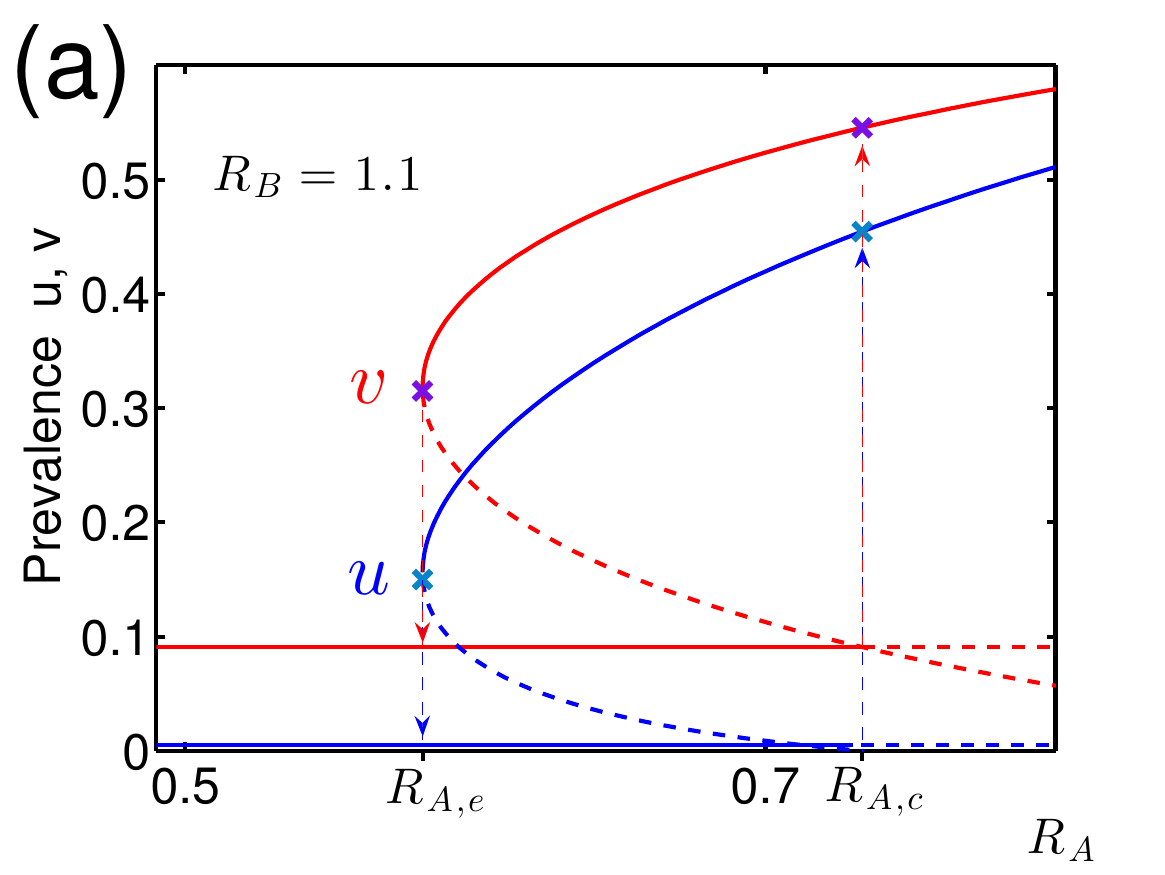}
\includegraphics[scale=0.45]{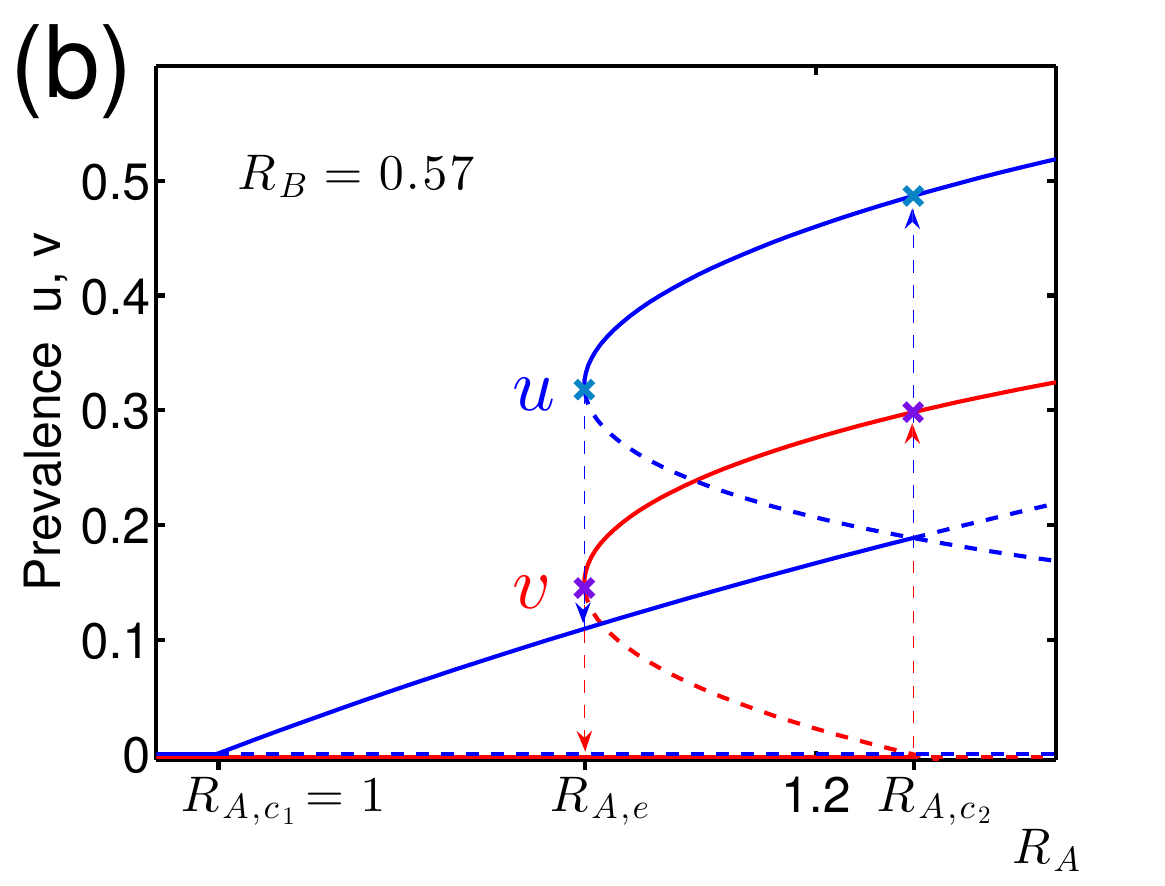}
\caption{\textbf{Bifurcations in asymmetric cooperative contagion}. The asymptotic prevalence $u^{\star}$ and $v^{\star}$ of agent $A$ and $B$, respectively, as a function of baseline reproduction ratio $R_{\text{A}}$ for fixed $R_{\text{B}}$ and cooperativity coefficient $\xi=5$, as indicated in Fig.~\ref{Fig:phasediagram}. (a) For $R_{\text{B}}=1.1$, a hysteresis structure emerges for agent $A$ between $u^{\star}=0$ and the $AB$ state branch, while for infection $B$ the hysteresis structure spans endemic state ($v^{\star}=1-1/R_{\text{B}}$) and the coinfection $AB$ branch. (b) For subcritical baseline $R_{\text{B}}=0.57$, prevalence of $A$ exhibits two outbreak transitions: i.) the classical transcritical bifurcation at $R_{A,c_{1}}=1$, ii.) a saddle node bifurcation with a hysteresis formed between an endemic ($u^{\star}=1-1/R_{A,c_{1}}$) and the coinfected branch within $R_{A,e}<R<R_{A,c_{2}}$ . Note that the second, discontinuous jump in $u$ when $R_{\text{A}}$ is increased beyond $R_{A,c_{2}}$ is caused because the state $v^{\star}=0$ loses stability at this point.
}
\label{Fig:cuts}
\end{figure*}
Because we focus on cooperative contagion we restrict the transmission rates: 
\begin{eqnarray}
\alpha_{\text{BA}}=\xi_{\text{B}}\alpha_{\text{A}}\geq\alpha_{\text{A}}, & \quad & \alpha_{\text{AB}}\equiv\xi_{\text{A}}\alpha_{\text{B}}\geq\alpha_{\text{B}},\label{eq:transmission_rates}
\end{eqnarray}
cooperativity thus implies that $\xi_{\text{A}},\xi_{\text{B}}\geq1$.
For example, a value $\xi_{\text{A}}=5$ means that transmission of
$B$ to an individual already carrying $A$ is 5-fold the transmission
compared to the baseline transmission to an $S$ individual. Based
on the above reactions one can obtain a set of ordinary differential equations
for the fraction of individuals in each state. The reactions above, however, suggest
a more suitable set of compartments $\mathcal{S}$, $\mathcal{A}^{+}$,
$\mathcal{B}^{+}$ and $\mathcal{AB}=\mathcal{A}^{+}\cap\mathcal{B}^{+}$
with the corresponding dynamical variables $s$, $u$, $v$, and $w$:
the fractions of susceptibles, individuals infected with $A$ (including
those that are also infected with $B$), individuals infected with
$B$ (including those that are infected with $A$), and individuals
infected with both $A$ and $B$, respectively, see Fig.~\ref{Fig:fig1}.
In the limit of a large, well-mixed host population the dynamics is
described by 
\begin{eqnarray}
\dot{u} & = & R_{\text{A}}su+\xi_{\text{B}}R_{\text{A}}(v-w)u-u\nonumber \\
\dot{v} & = & R_{\text{B}}sv+\xi_{\text{A}}R_{\text{B}}(u-w)v-v\nonumber \\
\dot{w} & = & \xi_{\text{A}}R_{\text{B}}(u-w)v+\xi_{\text{B}}R_{\text{A}}(v-w)u-2w\label{eq:dynsys_general},\\
s & = & 1-u-v+w,
\end{eqnarray}
where $R_{\text{A}}=\alpha_{\text{A}}/\beta$, $R_{\text{B}}=\alpha_{\text{B}}/\beta$
and time is measured in units of $\beta^{-1}$. For cooperativity
coefficients $\xi_{\text{A}}=\xi_{\text{B}}=1$ the above system describes
two independent contagion processes: If $R_{\text{A}},R_{\text{B}}>1$
the stable endemic state is given by $u^{\star}=1-R_{\text{A}}^{-1}$,
$v^{\star}=1-R_{\text{B}}^{-1}$, $w^{\star}=\left(1-R_{\text{A}}^{-1}\right)\left(1-R_{\text{B}}^{-1}\right)$,
and $s^{\star}=\left(R_{\text{A}}R_{\text{B}}\right)^{-1}.$

We now consider the effect of cooperativity. In the following and
in analogy to the labels used to identify the state of an individual
in the population, it is useful to assign the same labels $S$, $A$,
$B$, and $AB$ to the potential asymptotic states of the entire host
population. We say, e.g., that the system is in state $A$ if only
agent $A$ is present in the population, the contagion free state
is $S$, etc.. We begin with a symmetric system in which $\xi_{\text{A}}=\xi_{\text{B}}=\xi$
and identical baseline reproduction ratios $R_{\text{A}}=R_{\text{B}}=R$.
In this case the above system reduces to:
\begin{eqnarray}
\dot{u} & = & Rsu+\xi R(v-w)u-u\nonumber \\
\dot{v} & = & Rsv+\xi R(u-w)v-v\nonumber \\
\dot{w} & = & \xi R\left[2uv-(u+v)w\right]-2w.\label{eq:symmetricsystem}
\end{eqnarray}

Fig.~\ref{Fig:fig2} illustrates the bifurcation analysis of the system.
When $1\leq\xi<2$ the system exhibits a behavior similar to independent
contagion processes: At $R=1$ we observe a transcritical bifurcation
yielding a stable endemic population state $AB$ for $R>1$. This
means that even when cooperativity amplifies transmission rates by
up to a factor of 2, we see no qualitative dynamical difference. 

However, when cooperativity exceeds a critical magnitude, i.e. for
$\xi>\xi_{c}=2$, a different bifurcation behavior emerges. As $R$
is increased and before the conventional critical point $R_{c}=1$
is reached, a saddle-node bifurcation emerges at $R_{e}=2\sqrt{\xi-1}/\xi<1$.
When $R_{e}<R<R_{c}$, in addition to the trivial stable state $S$,
two $AB$ stationary states exist
\begin{eqnarray}
u_{\pm}^{\star}=v_{\pm}^{\star} & = & \frac{\xi R-2\pm\sqrt{\xi^{2}R^{2}-4\xi+4}}{2\xi R},\label{eq:knut}\\
w_{\pm}^{\star} & = & \frac{Ru_{\pm}^{\star}(\xi-2)+R-1}{(\xi-1)R},\nonumber 
\end{eqnarray}
one of which $(u_{+}^{\star},v_{+}^{\star},w_{+}^{\star})$ is stable,
the other unstable. Thus, when $R_{e}<R<R_{c}$ sufficiently small
perturbations to the $S$ state will have no effect as $S$ is stable.
However, when perturbations are sufficiently large, the system will
approach the endemic $AB$ state with $u_{+}^{\star}=v_{+}^{\star},w_{+}^{\star}$
. Furthermore, when $R$ is increased beyond the critical value $R_{c}=1$,
state $S$ loses stability and any arbitrarily small perturbation
will yield a discontinuous jump to the endemic state, reminiscent
of a first order phase-transition. For example when $R=R_{c}+\varepsilon$
with $0<\varepsilon\ll1$ the stable endemic state is $u_{+}^{\star}=v_{+}^{\star}\approx(\xi-2)/\xi$,
$w_{+}^{\star}\approx(\xi-2)^{2}/\xi(\xi-1)$. So if say $\xi=10$
this yields an endemic state in which $71\%$ of the population is
in state $AB$, immediately after $R_{c}$ is crossed. Cooperative
contagion also exhibits hysteresis: Starting with $R>R_{c}$ and state
$AB$, decreasing $R$ across the critical value $R_{c}$ from above
will not result in immediate extinction. A high endemic state is maintained
until the eradication threshold $R_{e}$ is reached, which can be
substantially smaller than the ordinary epidemic threshold $R_{c}$.
Decreasing $R$ below $R_{e}$ will then yield a sudden collapse of
$AB$ into $S$. 

\begin{figure*}[t!]
\includegraphics[scale=0.45]{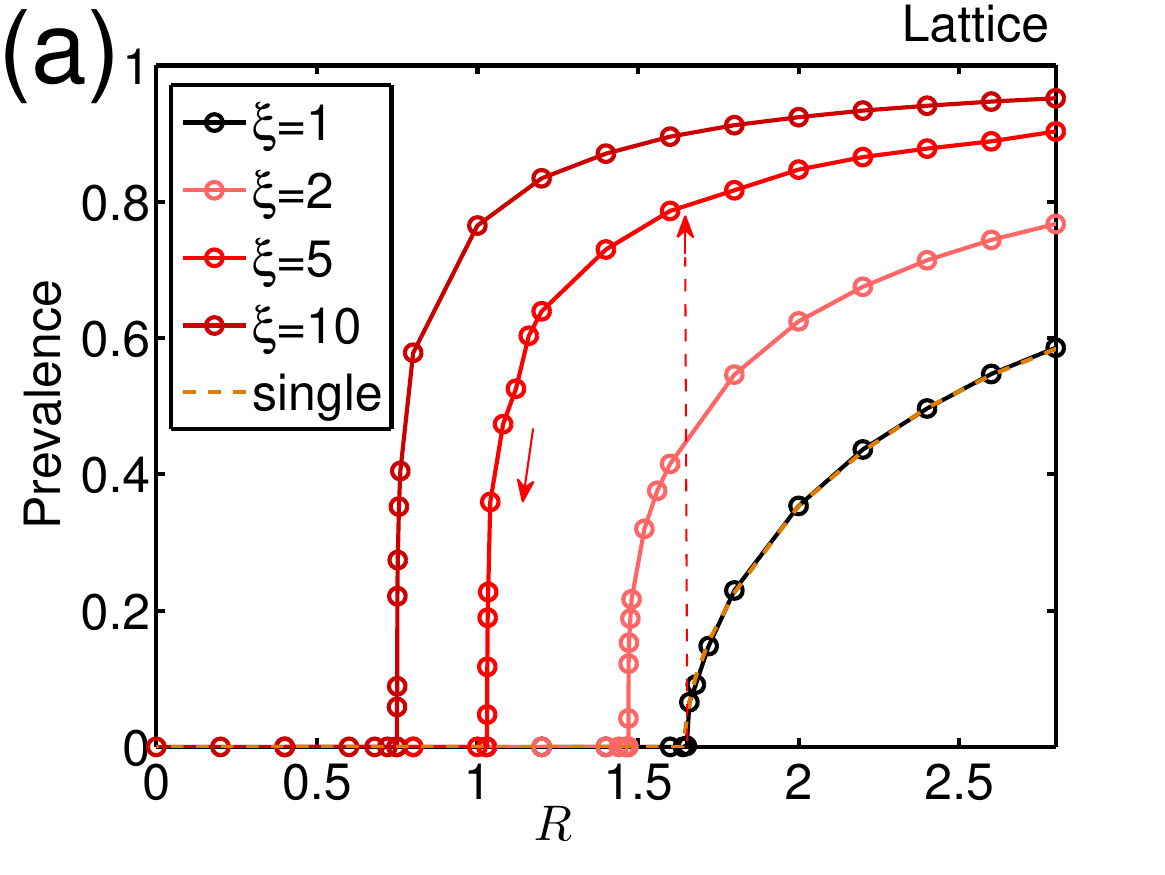}
\includegraphics[scale=0.45]{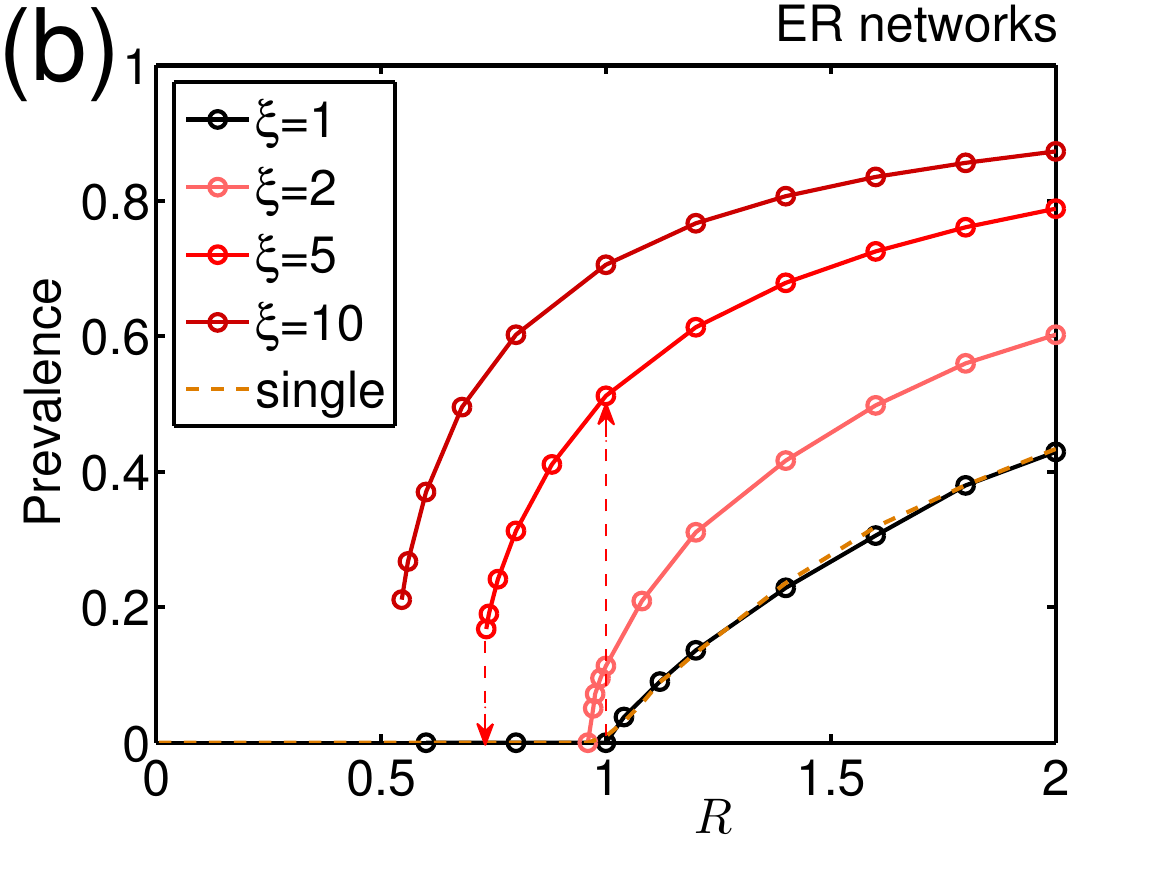}
\caption{\textbf{Phase transitions of prevalence (either $A$ or $B$) in stochastic network models}. Cooperative contagion on (a) two dimensional square lattices (with size $100\times100$) and (b) ER networks (with network size $N=10000$ and average degree $\langle k\rangle=4$). The reproduction ratio is defined as $R=\langle k\rangle\alpha/\beta$ where $\alpha$ is the transmission rate across a link. To investigate the extinction threshold, the simulations (with Gillespie algorithm) are initiated with complete prevalence. The transitions is obtained by decreasing $R$. Outbreak transitions are only possible when $R$ is close to the threshold of single infection if the population starts with tiny infected fraction, e.g. two remote infected nodes with $A$, $B$ respectively. The thresholds of single infection $R_{c}$ are around 1.64 and 1 respectively in (a) and (b), while a smaller eradication threshold is expected in strong cooperative cases as shown, therefore a hysteresis structure is formed in line with the above mean field theory. 
}
\label{Fig:perc}
\end{figure*}

Eqs.~(\ref{eq:symmetricsystem}) capture the symmetric special case
of the more general system defined by Eqs.~(\ref{eq:dynsys_general}),
the latter of which has four parameters, $R_{\text{A}}$, $R_{\text{B}}$,
$\xi_{\text{A}}$, and $\xi_{\text{B}}$. Fig.~\ref{Fig:phasediagram}
illustrates the phase diagram for a more general choice of these parameters.
Fixing the cooperativity coefficients to $\xi_{\text{A}}=\xi_{\text{B}}=5$
we investigated the phases of the system as a function of baseline
reproduction ratios $R_{\text{A}}$ and $R_{\text{B}}$. Apart from
the expected stable states we observe a rich variety of bistable states
in the region in which baseline reproduction is near unity, as is
illustrated in Fig.~\ref{Fig:cuts}. For example, when $R_{\text{B}}=1.1$
and starting with $R_{\text{A}}\approx0$ the system is initially
in state $B$. Increasing $R_{\text{A}}$ further a saddle-node bifurcation
occurs and the system enters a regime in which $B$ and $AB$ are
both stable. When $R_{\text{B}}<1$ , e.g. $R_{\text{B}}=0.57$, increasing
$R_{\text{A}}$ first yields an ordinary transcritical bifurcation
to the $A$ state, followed by a second bifurcation into a regime
in which $A$ and $AB$ are stable, and finally, a third bifurcation
to into the regime in which only $AB$ is stable, see also Fig.~\ref{Fig:phasediagram}.
A key property of the system is that the complexity of transitions
is only observed for baseline reproduction ratios near unity. If 
one baseline reproduction is too low or two high, only a single ordinary
transitions and no state coexistence is observed. This is an interesting
property from an evolutionary point of view. When new strains of transmissible
agents emerge, typically they are not adapted to the host and possess
baseline reproduction not significantly larger than unity, or even
smaller. Cooperativity with other transmissible agents and coexistence
of stable endemic states may present an opportunity for developing
a species rich system with higher evolvability. This type of complexity
is expected to increase dramatically when more than two transmissible
agents are involved, yielding a potentially rich space of stable states
and an increasing complexity in phase separation manifolds in parameter
space.
\begin{figure*}[t!]
\includegraphics[scale=0.6]{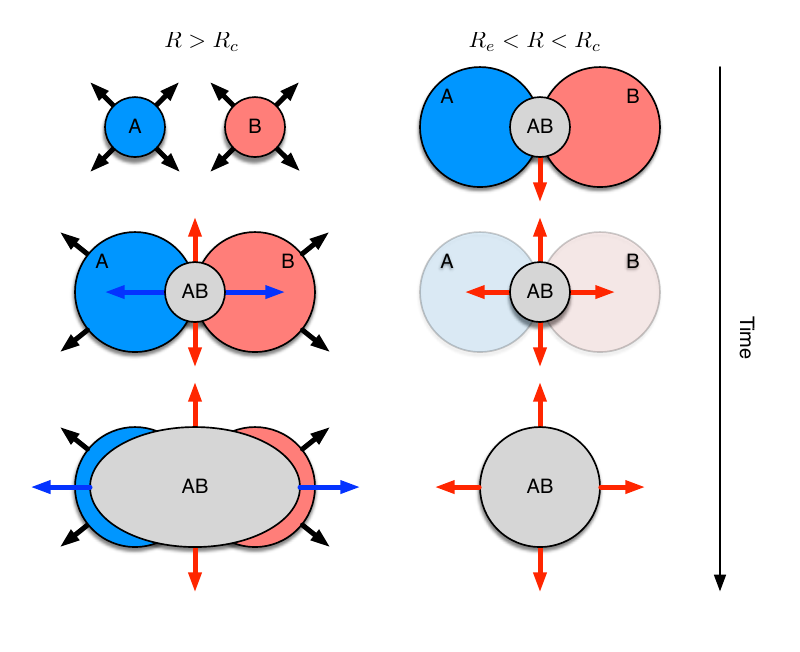}
\includegraphics[scale=0.45]{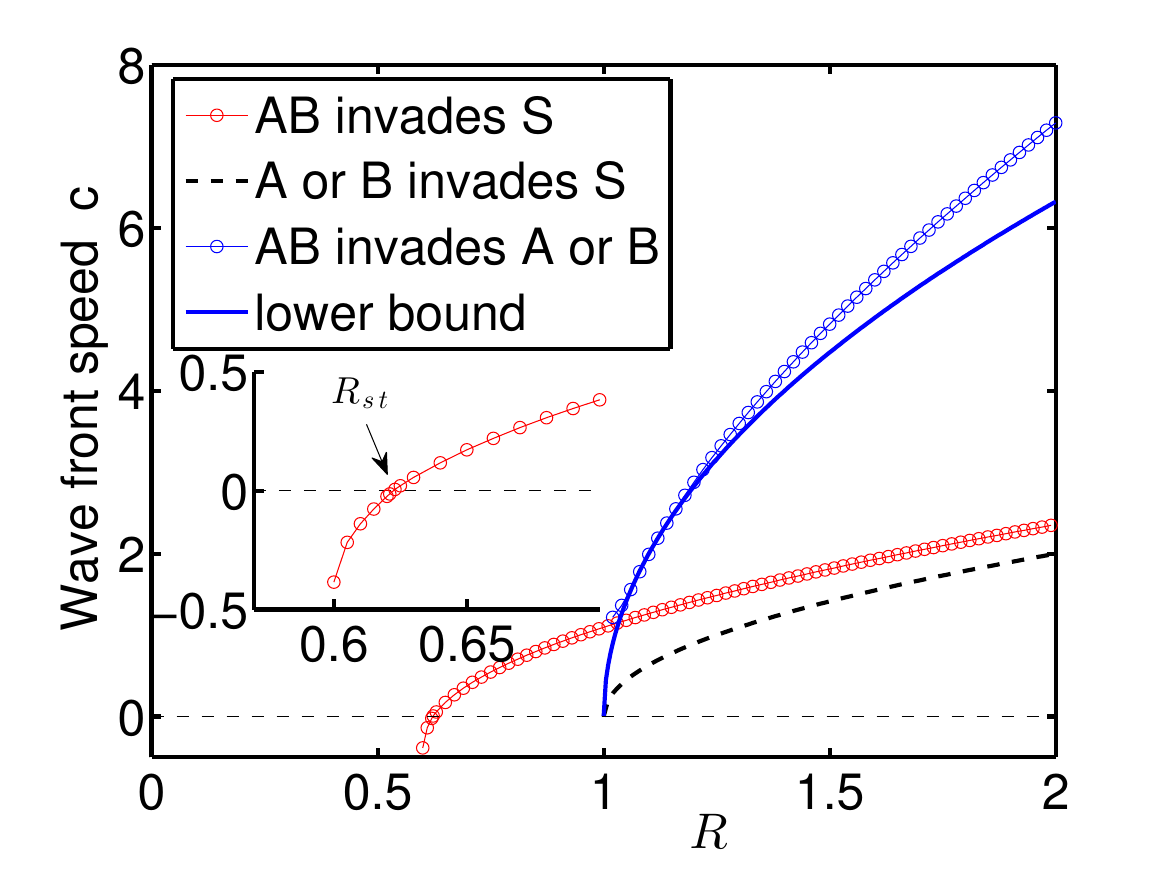}
\caption{\textbf{Spatial dynamics of cooperative contagion processes}. Left: When $R>R_{c}$ initially separated region of only $A$ or $B$ affected regions grow with a front speed equivalent to single SIS dynamics, $c_{0}\propto\sqrt{(R-1)D}$ (black arrows). When the wave fronts touch, a region of coinfection emerges (grey). This region touches the regions of susceptible (white background) and on each side region that are dominated by either $A$ or $B$ yielding two new front speeds associated with $AB$ invading $A$ or $B$ (blue arrow) and $AB$ invading $S$ (red arrow). Invasion of $AB$ into the susceptible region occurs at a speed $c_{AB\rightarrow S}>c_{0}$ the blue/red regions occurs at a speed $c_{AB\rightarrow A/B}>c_{AB\rightarrow S}$. In this transient phase, the patterns is shaped by three front speeds of different magnitude. This implies that the intermittent $AB$ invasion will take over the entire pattern and eventually only the $AB$ region will propagate into the susceptible region. When $R<R_{c}$ initially separated $A$ and $B$ regions cannot be sustained and will relax to the contagion free state. However, if initially a small overlap exists (a nucleation of $AB$) the pattern will eventually converge to $AB$ invading $S$ as well despite the fact that $R<R_{c}$. Right: Comparison of the three wave front speeds ($c_{AB\rightarrow S}$ – red circled line, $c_{A/B\rightarrow S}=c_{0}$ – black dashed, $c_{AB\rightarrow A/B}$ – blue circled line) as a function of $R$ in 1d space, together with a lower analytic bound for $c_{AB\rightarrow A/B}\approx\sqrt{\xi(R-1)D}=\sqrt{\xi}c_{0}$ (blue solid line). The inset shows that in the invasion of $AB$ into the susceptible region, the speed $c_{AB\rightarrow S}$ is positive for $R>R_{st}$, and negative for $R_{e}<R<R_{st}$, with $R_{st}=0.62218(4)$ and $R_{e}=0.6$. Parameters: $\xi=10$, $D=1$. 
}
\label{Fig:speeds}
\end{figure*}

The deterministic model discussed above cannot account for fluctuations
or population heterogeneities. An important question is therefore
whether the observed phenomena prevail in a more complex scenario
in which transmissions and recovery events are stochastic, the host
population is finite and not every host interacts with every other
host at equal rates. Typically, stochastic effects in a well-mixed
system are modeled by birth-death type stochastic processes equivalent
to the reactions depicted in Fig.~\ref{Fig:fig1} and generating solutions
to the corresponding master-equation for a fixed but finite population
size $N$. Population heterogeneities are typically addressed by modeling
these processes on fixed network topologies or lattices in which host
individuals only interact with the neighbors defined by the network
links. In order to address the robustness of effects and properties
derived for the deterministic system of Eqs.~(\ref{eq:dynsys_general})
we investigated cooperative contagion dynamics in a stochastic 2d-lattice
system and and Erd\H{o}s-R\'enyi (ER) network with equal mean degree
and number of nodes. The results are compiled in Fig.~\ref{Fig:perc}. In both cases,
we observe hysteresis, and a separation of extinction and outbreak
thresholds for large cooperativity coefficients $\xi$. Interestingly,
the extinction transition is continuous in the lattice, a consequence
of the local coupling of the system. The ER network exhibits discontinuous
transitions, as predicted by the above mean field treatment. This
is not surprising as the ER network is topologically more similar
to the well-mixed scenario. Based on these observations, we believe
that the key features of cooperative contagion can be expected also
in more realistic, structured populations \cite{SM1}.

\section{Wave propagation in spatially extended systems}
\label{Sec:wave}
An important aspect of contagion processes is their spatial propagation.
When simple contagion processes with $R>1$ are seeded in a spatially
homogeneous susceptible host population and contagion dynamics is
combined with diffusive dispersal of the host these systems typically
exhibit propagating wavefronts that travel at constant speeds. The
endemic state invades the unstable $S$ domain. The generic SIS contagion
process, e.g., can be described by
\begin{equation}
\partial_{t}u=R(1-u)u-u+D\partial_{x}^{2}u,\label{eq:spatialSIS}
\end{equation}
where $u=u(\mathbf{x},t)$ is the density of infected individuals
at location $\mathbf{x}$ at time $t$. The combination of local initial
exponential growth (for $R>1$) and diffusion yields a front-speed
depending on the basic reproduction ratio and diffusion coefficient
$D$: $c=2\sqrt{(R-1)D}$. This is a generic feature of processes
that exhibit pulled fronts~\cite{murray2003}. Given the more complex
nature of cooperative coinfection, especially the dynamical bistability
for intermediate baseline reproduction ratio $R_{e}<R<R_{c}$ and
large cooperativity coefficient $\xi$, we can expect a richer set
of phenomena when cooperative contagion processes expand in space.
To account for a diffusing host we extend Eqs.~(\ref{eq:dynsys_general})
and consider the corresponding reaction-diffusion system:
\begin{eqnarray}
\partial_{t}u & = & f_{u}(u,v,w)+D\partial_{x}^{2}u\nonumber \\
\partial_{t}v & = & f_{v}(u,v,w)+D\partial_{x}^{2}v\nonumber \\
\partial_{t}w & = & f_{w}(u,v,w)+D\partial_{x}^{2}w\label{eq:spatial}
\end{eqnarray}
where $D$ in last terms is the diffusion coefficient and the functions
$f_{u}$, $f_{v}$, and $f_{w}$ are the same as on the rhs. of Eqs.~(\ref{eq:symmetricsystem}).
The dynamical variables are function of time $t$ and space $\mathbf{x}$,
e.g. $u=u(\mathbf{x},t)$. We assume that the diffusion coefficient
is constant and independent of the state of a host individual primarily
focusing on contagion processes that do not affect the host's dispersal
behavior. We also consider a constant overall density which implies
that $s=1-u-v+w$ at every position $\mathbf{x}$. As before we use
labels $S$, $A$, $B$ and $AB$ to refer to region that are contagion
free, only affected by $A$, only affected by $B$, and both $A$
and $B$, respectively.
\begin{figure*}[t!]
\includegraphics[scale=0.47]{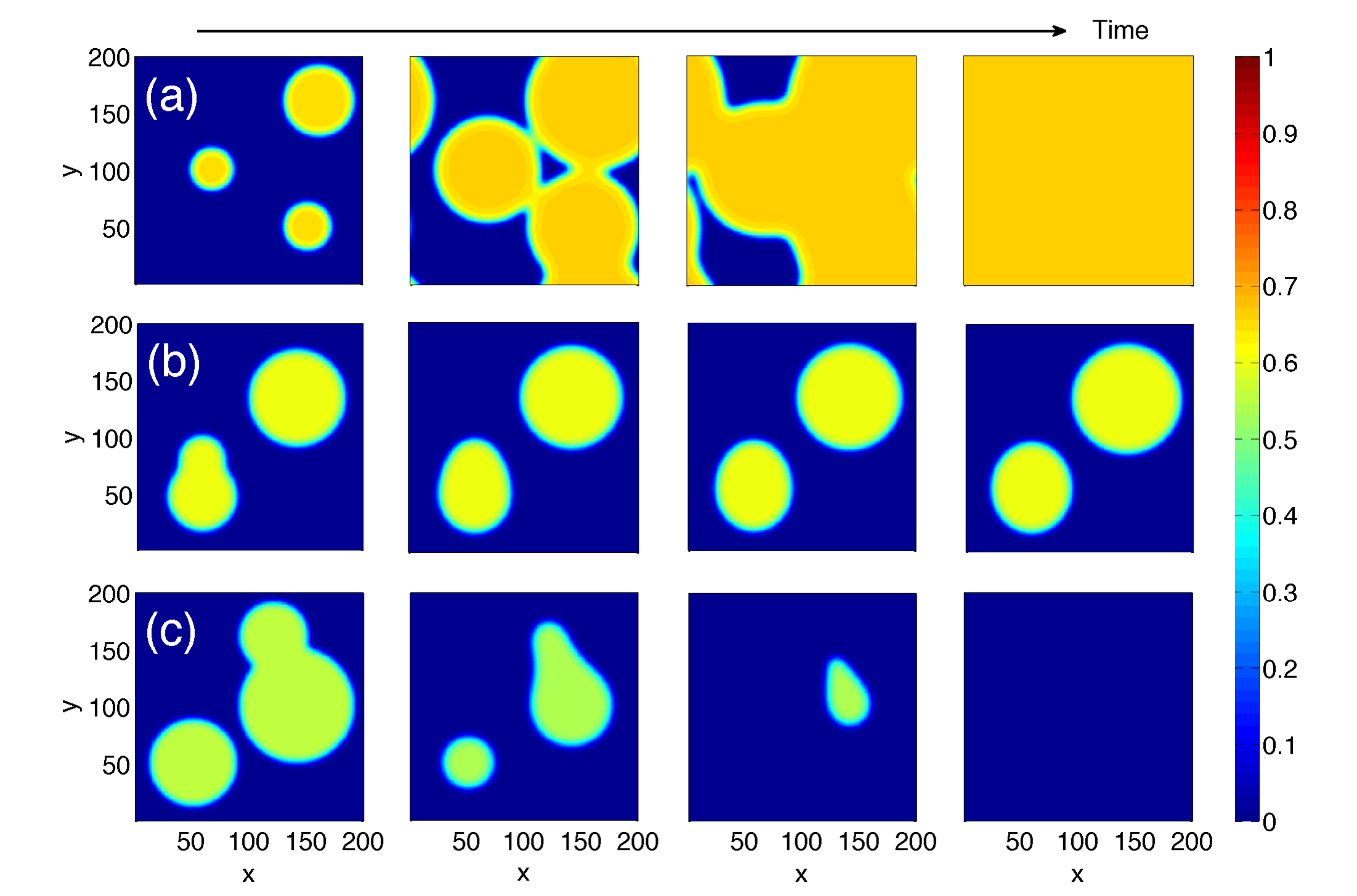}
\caption{\textbf{Three typical contagion propagation modes in 2d space}. Top panels (a): forward propagation ($R=0.65$). Middle panels (b): standing front ($R=R_{st}=0.62671(4)$). Bottom panels (c): backward propagation ($R=0.61$). The infected fraction $1-s$ is color coded. The sequence (from left to right) of panels depicts the time course of the infected regions at time t=0, 100, 200, 300, respectively. Here $R_{st}$ is slightly different from the value in 1d space, up to the dimension correction. Initial conditions start from some randomly infected round regions with random radius as shown in the first column. Other parameters: $D=1$, $\xi=10$, where $R_{e}=0.6$.
}
\label{Fig:fig7}
\end{figure*}

The system defined by Eqs.~(\ref{eq:spatial}) exhibits a range of
front velocities, each one corresponding to different states invading
regions in a different state. For example a localized $A$-patch invades
an $S$-region at a different speed than a uniform $B$-region (turning
the latter into an $AB-$region). A localized $AB$-patch invades
an $S$-region differently than an $A$-region. To understand the
asymptotics and transients of the system we first consider a uniform
population in state $S$, with the exception of two localized patches,
each being in state $A$ and $B$ respectively and separated by some
distance, see Fig.~\ref{Fig:speeds}. When $R>R_{c}$ cooperative
contagion plays no role at the beginning, each patch will expand at
a constant front speed of $c_{0}\propto\sqrt{(R-1)D}$. Once these
growing patches touch, cooperativity kicks in at the $A$-$B$ interface.
The emerging $AB$-nucleus has interfaces to the $A$ and $B$ regions
as well as to the $S$ region. For $\xi>1$ the invasion of $AB$
into the $S$-region is faster at a speed $c_{AB\rightarrow S}>c_{0}$
as expected. Interestingly, the invasion of the $AB-$region into
$A$-region (and $B$-region) occurs at an even higher speeds $c_{AB\rightarrow A,B}>c_{AB\rightarrow S}$.
Using a propagating wave ansatz $u=u(x-ct)$ for a 1-d spatial support
(analogously for variables $v$ and $w$) one can compute a lower
bound $c_{AB\rightarrow A,B}\approx\sqrt{\xi(R-1)D}=\sqrt{\xi}c_{0}$.
Because $c_{AB\rightarrow A,B}>c_{AB\rightarrow S}$ the system will
eventually converge to a uniform $AB$-region that spreads at speed
$c_{AB\rightarrow S}$. Regions affected only by one agent will not
persist. This effect of enhanced wave-front speed might be particularly
relevant in situations in which a covert, unknown and commensal agent
is endemic in some region and a known process with known baseline
reproduction ratio expands somewhere else in the system at a speed
that is computed based on its baseline reproduction ratio. If this
front enters the region in which the unknown but highly cooperative
covert agent prevails, a sudden but potentially unexpected boost in
the proliferation of the initial spreading process could occur. 

In the bistable region $R_{e}<R<R_{c}$, isolated islands of \emph{A}
nor \emph{B} cannot persist. If we initialize the system with $A$
and $B$ patches that share a small overlapping region in the $AB$
state cooperativity can yield the survival of the $AB$ state while
the homogeneous $A$ and $B$ states fade. The remaining $AB$ patch
then proliferates at speed $c_{AB\rightarrow S}$. Interestingly,
we observe negative propagation speed in this regime, $c_{AB\rightarrow S}<0$,
which implies a receding \emph{AB}-region. This behavior is caused
by the dispersal of $A$ or $B$ affected individuals into the $S$-region
in which the $S$-state is also stable. Once individuals enter this
state, they have a higher likelihood of becoming susceptible than
being colonized by both agents. The wavefront acts as a drain for
infected agents and a competition exists between the supply of new
agents of type $A$ or $B$ and the diffusive dilution of their concentration.
For a critical choice of parameters, e.g. the baseline reproduction
ratio we observe a stationary heterogeneous solution, an immobile
front, that separates $S$ from $AB$ regions. Figure \ref{Fig:fig7}
illustrating the three typical propagation modes in 2d space are depicted
(see also the movies in the Supplemental Materials \cite{SM2}).

\section{Discussion}
\label{Sec:Discussion}
We present a reaction kinetic model for the dynamics of cooperative contagion of the susceptible-infected-susceptible class. The most prominent property of the model is the existence of discontinuous transitions to an endemic state when the traditional outbreak threshold is crossed and a separation of outbreak and extinction thresholds, the magnitude of which depends on the degree of cooperativity. Although we derive the key properties analytically and numerically in a deterministic model suitable for large, well-mixed populations, we observe the key features of discontinuous transitions also in a stochastic network variant of the model. The system of two cooperating agents that proliferate in a host populations exhibits diverse properties when spatial diffusion is incorporated yielding different types of transients and spreading speeds. 

Although we discussed the model predominantly in the context of the spread of transmissible diseases, the model is sufficiently generic to be applied to other transmissible contagion processes that influence each other cooperatively, e.g. the adoption of one technology may increase the “infection” of a user with another type of technology which may then occur explosively or at different speeds than expected. 

Considering a model for only two interacting agents is foundation that can easily be generalized to a larger number, potentially a network, of interacting agents. If the baseline reproduction of a family of transmissible agents is in the critical regime, we expect in such a system a diverse set of stable configurations and we believe that the model presented here is a helpful starting point for investigating these more general systems.

\section*{Acknowledgement}
L.C. and F.Gh. would like to thank W. Cai and P. Grassberger for valuable comments and discussions. D.B. acknowledges fruitful discussion with Osamah Hamouda and Imbish Mortimer. L.C. also would like to thank the hospitality of Beijing Computational Science Research Center (CSRC) for a pleasant stay during the final revision stage.

Author contributions: L.C., F.Gh, D.B. conceived research, L.C. and D.B. developed dynamical model, L.C., D.B. theoretical, analytical treatment of mean field and spatial model, L.C. performed numerical analysis. D.B. and L.C. wrote manuscript.


\end{document}